\documentclass[pra,aps,10pt,twocolumn,superscriptaddress,longbibliography]{revtex4-1}
\usepackage{amsmath}
\usepackage{amssymb}
\usepackage{amsthm}
\usepackage{amsfonts}
\usepackage{enumerate}
\usepackage{latexsym}
\usepackage{bm}
\usepackage{graphicx}
\usepackage{subfigure}
\usepackage{color}
\usepackage[dvipsnames]{xcolor}
\usepackage[colorlinks]{hyperref}
\usepackage{ulem}
\usepackage{cancel}

\newcommand{\beq}{\begin{equation}}
\newcommand{\eneq}{\end{equation}}

\input{epsf}

\begin{document}

\title{Interplay of vibration- and environment-assisted energy transfer}

\author{Zeng-Zhao Li} 
\affiliation{Department of Chemistry, University of California, Berkeley, California 94720, USA}
\author{Liwen Ko} 
\affiliation{Department of Chemistry, University of California, Berkeley, California 94720, USA}
\author{Zhibo Yang} 
\affiliation{Department of Chemistry, University of California, Berkeley, California 94720, USA}
\author{Mohan Sarovar}
\affiliation{Sandia National Laboratories, Livermore, CA, 94551, USA}
\author{K. Birgitta Whaley}
\affiliation{Department of Chemistry, University of California, Berkeley, California 94720, USA}
\affiliation{Berkeley Center for Quantum Information and Computation, Berkeley, California 94720, USA}

\begin{abstract}
We study the interplay between two environmental influences on excited state energy transfer in photosynthetic light harvesting complexes, namely, vibrationally assisted energy transfer (VAET) and environment-assisted quantum transport (ENAQT), considering a dimeric chromophore donor-acceptor model as a prototype for larger  systems. We demonstrate how the basic features of the excitonic energy transfer are influenced by these two environments, both separately and together, with the environment being fully quantum in the case of VAET and treated in the Haken-Strobl-Reineker classical limit in the case of ENAQT. Our results reveal that in the weak noise regime, the presence of a classical noise source is detrimental to the energy transfer that is resonantly assisted by the exciton-vibration interactions intrinsic to VAET. In the strong noise regime we reproduce all the features of ENAQT including the turnover into a Zeno regime where energy transfer is suppressed, and VAET is insignificant.
\end{abstract}
\date{\today}
\maketitle

\section{Introduction}

Recent years have seen a rapid growth in understanding how underdamped vibrational degrees of freedom play a role in efficient exciton transport in photosynthetic light harvesting systems \cite{chinRoleNonequilibriumVibrational2013,Irish14pra,Sato14jcp,KilloranHuelgaPlenio15jcp,Lee17jcp,KolliOlaya12jcp,Juhasz18aip,NalbachThorwart15pre,OReillyOlaya14ncomms}, and in the long time room-temperature coherence observed in these systems \cite{CinaFleming04jpca,Christensson12jpcb,PlenioHuelga13jcp,Scholes17nature,WangEngel19natrevchem}. This phenomenon has been referred to as vibrationally-assisted energy transfer (VAET) in the literature~\cite{GormanSarovarHaeffner18prx}. 
At the same time, fluctuating environments, such as those due to overdamped vibrational modes, have also been shown to promote exciton transport under certain conditions, a phenomenon termed environment-assisted quantum transport (ENAQT)~\cite{RebentrostAspuru09njp,ChenZhao15}. In particular, the latter occurs when the excitonic degrees of freedom interact with the environment via a pure dephasing interaction, which can be understood as the direct manifestation of a quantum random walk with dephasing~\cite{Kendon_2007}.

In this work we study the interplay between these two environment-driven exciton transport mechanisms. Earlier theoretical work has established that it is possible to observe  oscillations due to excitonic-vibronic quantum coherence even in the presence of additional strong environmental noise~\cite{DijkstraCaoFleming15jpcl}.
We extend this line of study here by analyzing a dimeric chromophore donor-acceptor system to provide a comprehensive picture of the regimes where either VAET or ENAQT dominate, 
as well as the effect of the interplay between these two processes on the energy transfer efficiency in intermediate regimes.
VAET is modeled by explicit treatment of an underdamped vibrational mode, while the environmental effect of the overdamped modes is captured by the addition of a classical noise, pure dephasing process (the Haken-Strobl-Reineker model \cite{Capek_1993}). While all the details of the vibrational environment of natural photosynthetic systems are not captured by such a classical noise model -- most importantly, thermal effects and relaxation are not captured -- the phenomenon of ENAQT has been observed within such a model of environmental noise \cite{RebentrostAspuru09njp}, and therefore it is sufficient in the first instance to study the interplay between this phenomenon and VAET. 

The model we study is particularly suited to experimental validation by trapped ion quantum simulators which could also simulate other interesting physics such as Dirac dynamics~\cite{SongMuga17pra}. This platform has been used to experimentally study VAET \cite{GormanSarovarHaeffner18prx}, and the addition of classical fluctuations can be easily achieved by noisy modulation of addressing lasers. We have also recently studied the effect of multiple underdamped vibrational modes on energy transfer within the context of a trapped ion simulation \cite{LiSarovarWhaley20}. We expect that the predictions made below for regimes of energy transfer can also be validated on the circuit-QED platform \cite{Potocnik_2018}.

The remainder of the paper is organized as follows. In Sec. \ref{sec:model} we summarize the model 
and our theoretical approach. In Sec. \ref{sec:invariance} we present an analysis of the symmetries in our model that explain the physical relevance of some of the parameters, especially in the trapped ion simulation of the model. In Sec. \ref{sec:features} we undertake numerical simulations to analyze 
various aspects of the interplay of VAET and ENAQT. Specifically, we find that the presence of additional classical dephasing noise tends to weaken the effects of vibrational assistance of energy transfer, with the VAET peaks eventually disappearing as the variance of the classical noise is increased. 
The optimum transfer efficiency is obtained at zero classical noise, where the only environmental effect is vibrational assistance of energy transfer from a vibration that is resonant with an excitonic energy difference. At larger values of classical noise, the energy transfer efficiency can be maximized at a finite noise variance and the resulting dephasing-enhanced energy transfer is found to occur at a given finite dimer energy gap that could be either resonant or off-resonant with the vibrational mode. Finally, in Sec. \ref{sec:concl} we conclude with a brief discussion.

\section{A dimeric noisy chromophore donor-acceptor system}
\label{sec:model}

\begin{figure}
\centering
  \includegraphics[width=.98\columnwidth]{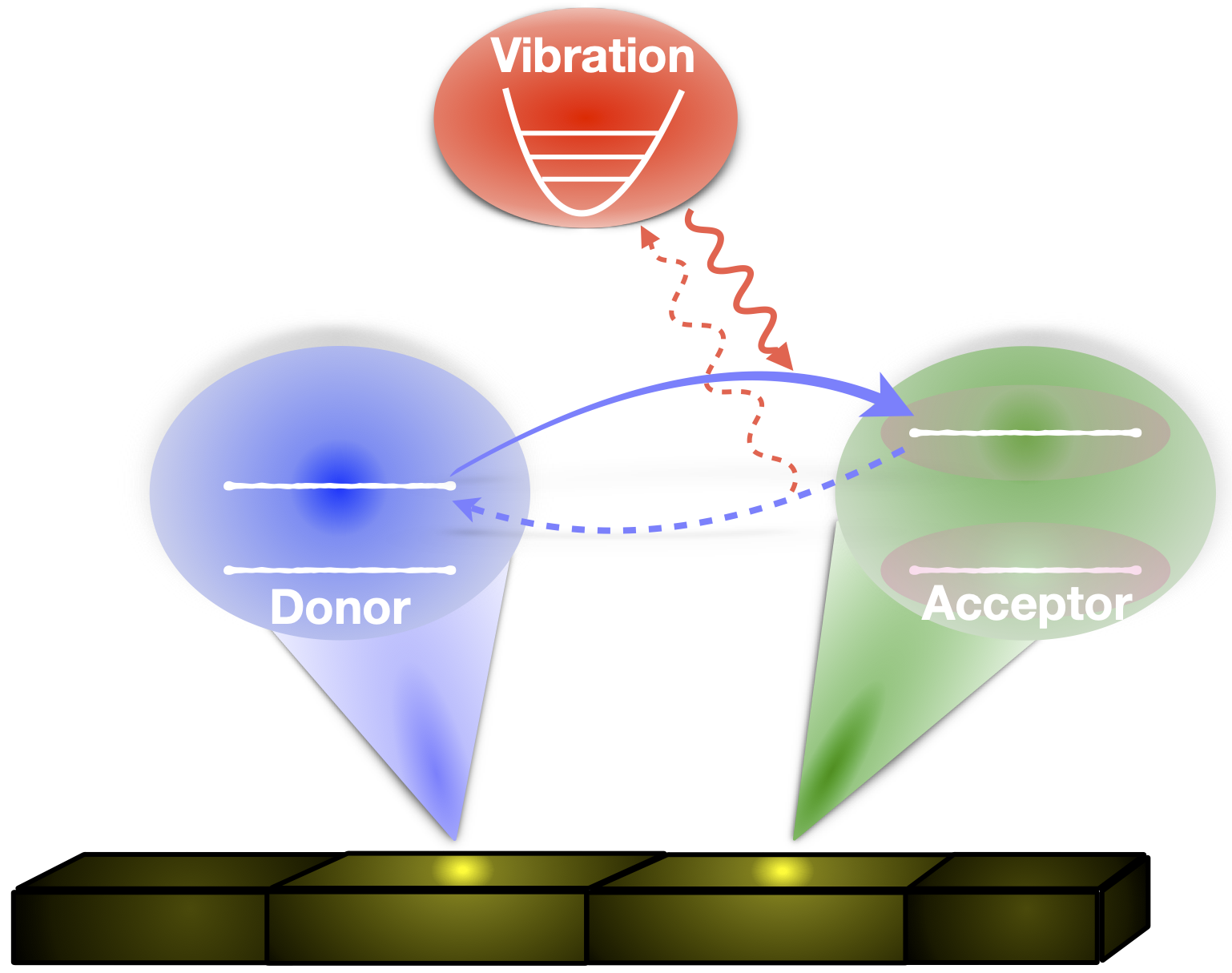} 
\caption{(color online) A schematic diagram of the dimeric noisy chromophore donor-acceptor system as simulated on a trapped-ion platform. The yellow ions in the chain represent the donor and acceptor species. The magenta ellipse  encircling the energy levels of the acceptor indicates the energetic shifts induced by classical stochastic fluctuations of the environment. The basic mechanism of excitation energy transfer from the donor to the acceptor (blue sold arrow) may be assisted by quantum noise in a form of the single vibration (red solid arrow). Dashed arrows illustrate the inverse downhill process.}
\label{fig:noisyVAETmodel}
\end{figure}

A basic model for demonstrating the VAET processes is a dimeric chromophore donor-acceptor system~\cite{GormanSarovarHaeffner18prx}. We therefore exploit it here as a prototype for larger photosynthetic energy transfer systems to explore the interplay between quantum and classical noise. 
The noisy VAET system is schematized in Fig.~\ref{fig:noisyVAETmodel} and is described by the Hamiltonian
\begin{eqnarray} \label{eq:Hamiltonian}
H&=&\frac{1}{2}\omega_d\sigma_z^{(d)}+\frac{1}{2}[\omega_a-\delta(t)] \sigma_z^{(a)}+\frac{1}{2}J\sigma_x^{(d)}\sigma_x^{(a)} \notag\\
&&+\nu a^{\dagger}a+\frac{1}{2}\kappa\sigma_z^{(a)}(a+a^{\dagger}),
\end{eqnarray}
where $\sigma_z^{(i)} = |e\rangle_i\langle e| - |g\rangle_i\langle g|$ and $\sigma_x^{(i)} = |e\rangle_i\langle g| + |g\rangle_i\langle e|$ with $i=d,a$. 
Eq.~\eqref{eq:Hamiltonian} includes donor (d) and acceptor (a) sites, each of which is modeled by a two-level system with transition frequency $\omega_i$, as well as the excitonic coupling $J$ between these. The single vibration denoted by annihilation/creation operators $a/a^{\dagger}$ coupled to the acceptor is a source of quantum noise and may coherently assist the excitation energy transfer~\cite{GormanSarovarHaeffner18prx}. 
The term $\delta(t)$ added to the site energy describes a classical Gaussian white noise source within the Haken-Strobl-Reinker model~\cite{Capek_1993, hoyer2010, SarovarWhaley11pre, FujitaAspuru14jcp}, 
i.e., $\delta$ at any time instant is distributed as $1/\sqrt{2\pi\sigma^2}e^{-\delta^2/(2\sigma^2)}$, characterized by zero mean and variance $\sigma^2$.
Here, we consider only diagonal fluctuations that are typically larger than fluctuations of the inter-molecular couplings~\cite{Cho_2005,Adolphs_2006} and the decoherence is dominated by pure dephasing~\cite{Capek_1993}. 

We focus here on the single-excitation manifold. This subspace is spanned by the basis states $|eg\rangle$ and $|ge\rangle$. With the projection operator $\Pi=|eg\rangle\langle eg| + |ge\rangle\langle ge|$, we obtain an effective Hamiltonian
\begin{eqnarray}
\tilde{H}&=&\Pi H\Pi = \frac{1}{2}[\Delta+\delta(t)]\tilde{\sigma}_z + \frac{1}{2}J\tilde{\sigma}_x \notag\\
&&+ \nu a^{\dagger}a +  \frac{1}{2}\kappa\tilde{\sigma}_z(a+a^{\dagger})
\label{eq:Hamiltonian_SingleEx}. 
\end{eqnarray}
Here $\Delta(=\omega_d-\omega_a)$ is the difference between the excitation energies of donor (d) and acceptor (a). The Pauli operators in this two-dimensional single-excitation subspace are defined as $\tilde{\sigma}_z=|eg\rangle\langle eg| - |ge\rangle\langle ge|$ and $\tilde{\sigma}_x=|eg\rangle\langle ge| + |ge\rangle\langle eg|$. 
In the absence of noise, negative $\Delta$ means uphill energy transfer from the donor to the acceptor that may be accompanied by an absorption of a phonon from the the vibrational mode to assist the transfer, while positive $\Delta$ corresponds to downhill transfer that can be enhanced by a phonon emitted to the vibration~\cite{GormanSarovarHaeffner18prx}. 
When the classical noise is present, the energy levels fluctuate. The form of the coupling term $\tilde{\sigma}_z$ implies that these two excitations are anticorrelated via the noise source~\cite{Uchiyama18njpqi,LiSarovarWhaley20}. This means that the noise can bring the excitation levels to resonance and thereby enhance the excitation energy transfer efficiency. 

The chromophore donor-acceptor dimeric system shown in Fig.~\ref{fig:noisyVAETmodel} has been experimentally engineered on a trapped-ion platform~\cite{GormanSarovarHaeffner18prx}. The energy sites can be encoded in internal electronic state of the ions, for example, ${\rm Ca}^+$ ($|S\rangle$ ($m_j = 1/2$) and $|D\rangle$ ($m_j = 1/2$)) and therefore the single excitation states are represented by the combined state (e.g., $|DS\rangle$ and $|SD\rangle$).
The interaction between the sites can be engineered via a bichromatic laser beam along the axis of the trap to be a two-qubit M$\o{}$lmer-S$\o{}$rensen quantum interaction and the site-vibration coupling can be achieved via a tightly focused laser beam localized to each ion. 
The dephasing noise represented by $\delta(t)\sigma_z^{(a)}/2$ causing instability of the acceptor splitting in Eq.~(\ref{eq:Hamiltonian}) might be incorporated in the trapped-ion platform by either engineering fluctuations of local magnetic fields acting on internal electronic states of the ion~\cite{Schindler_2013} or modulating a Stark-shift generated by a laser beam. 
In Appendix~\ref{app:technique}, we present a microscopic derivation of the site-vibration coupling that allows a direct mapping between the Hamiltonian of molecular photosynthetic systems and an emulation of the Hamiltonian on platforms such as trapped ions. 
{\color{red}}

To demonstrate how the VAET process is influenced by the classical noise, we consider the physical quantities
\begin{eqnarray} \label{eq:Pa}
P_a&=&\frac{1}{N_r}\sum_{i=1}^{N_r} P_{a,i}, \\
\eta_a&=&\frac{1}{N_r}\sum_{i=1}^{N_r} \eta_{a,i} ,
\end{eqnarray}
where $P_{a,i}(t)=Tr(\Pi_a\rho_i(t))$ and $\eta_{a,i}=\frac{1}{t_f}\int_0^{t_f} P_{a,i}(t) dt$ are the transfer probability and efficiency, where the latter is defined as the accumulated acceptor population during a given time period $t_f$, for each noise realization. Here $\Pi_a=|ge\rangle\langle ge|$ is the projection operator onto the excited state at acceptor site $a$ and $N_{r}$ is the number of noise realizations over which we average. 
The total density matrix operator for the $i$th noise realization is $\rho_i(t)=U\rho_s(t=0)\rho_bU^{\dagger}$, with $U=e^{-i\tilde{H}t}$.  
For the calculations shown here the initial states are $\rho_s(t=0)=|eg\rangle\langle eg|$ and $\rho_b=\sum_{n=0}^{\infty} \frac{n_b^n}{(n_b+1)^{n+1}} |n\rangle\langle n|$ for the donor-acceptor dimer and for the vibration, respectively. 
In our work the temperature of the vibration is quantified by the average phonon number $n_b$ via the relation $n_b=\frac{1}{e^{h\nu/k_{B}T}-1}$ with $h$ and $k_B$ being the Planck and Boltzmann constants, respectively, implying more phonons at a higher temperature.
The number of the noise realizations and Fock space size of the vibration are made large enough to make sure the accuracy and convergence of our results, i.e., $N_r=800$ and $N=24$.
In the numerical simulations below we use values for parameters that are typical of trapped-ion energy scales (i.e., kHz frequencies), rather than values typical of natural photosynthetic systems. However, these can be related by a simple scaling of energies (see, e.g., Table I in Ref. \cite{LiSarovarWhaley20}).

\section{An invariance of the VAET system \label{sec:invariance}}

In this section, we show that the effective Hamiltonian in Eq. (\ref{eq:Hamiltonian_SingleEx}) possesses an important symmetry implying that, for the quantities we compute here, only the difference in sign between the excitation energy difference $\Delta$ and the vibrational frequency $\nu$ is significant. This symmetry has some practical importance in the context of trapped-ion simulation of these dynamics since sweeping $\nu$ to negative values can be easier than sweeping $\Delta$ to negative values. This is because the former is defined by a difference in frequencies that can be tuned,  while the latter corresponds to an energy gap, which is more difficult to tune~\cite{GormanSarovarHaeffner18prx}.

\begin{figure}
\centering
  \includegraphics[width=.99\columnwidth]{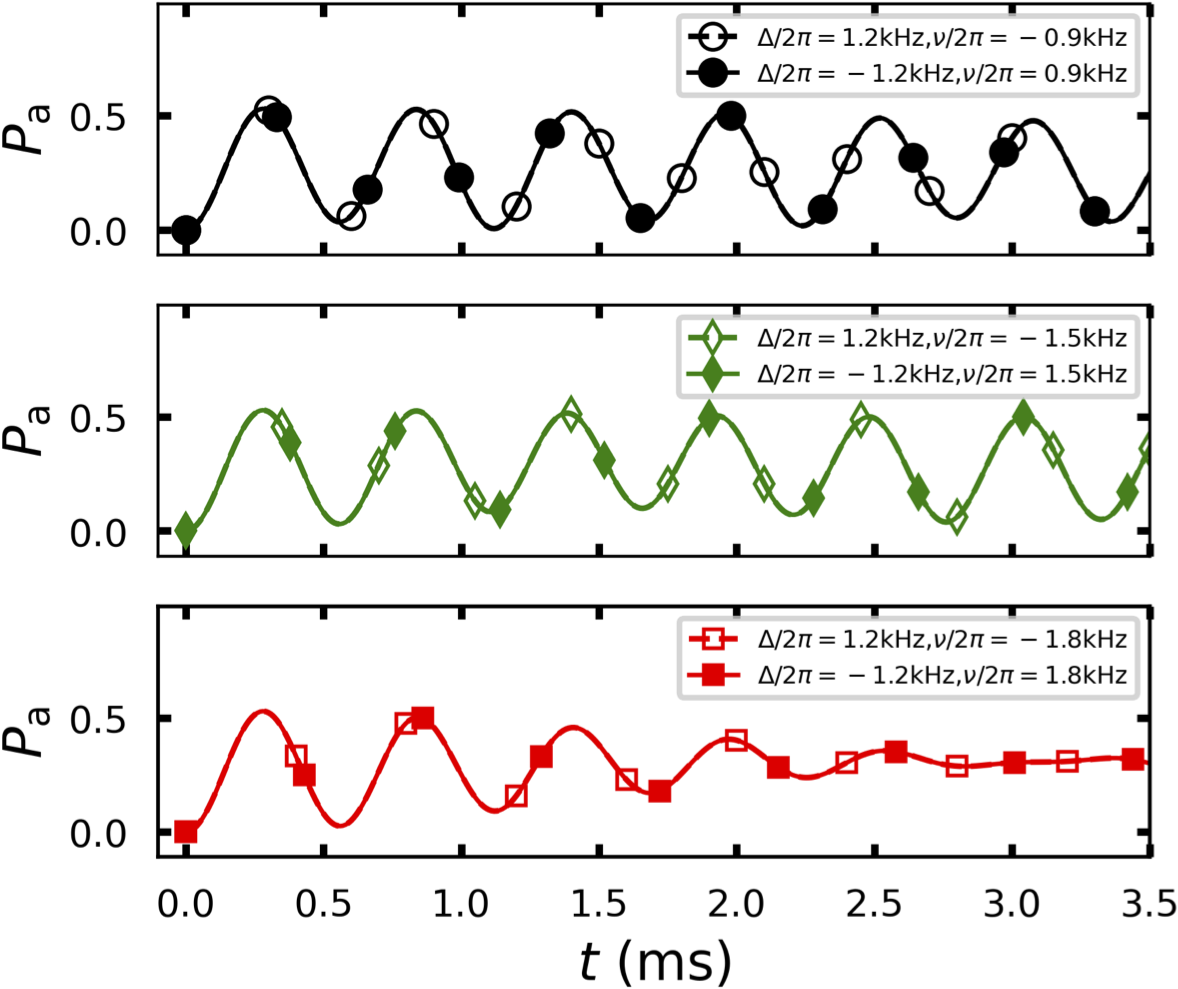} 
\caption{(color online) An example of the invariance of the transfer probability $P_a$ under the simultaneous sign change of $\Delta$ and $\nu$ in the dimeric VAET system. The red curves denote the resonant case ($\Delta^2+J^2=\nu^2$) while two other curves refer to off-resonant cases. We take $J/2\pi=1.3$kHz, $\kappa/2\pi=0.229$kHz, $n_b=0.4$, and $N_r=800$. The classical noise $\delta =0$ is used in these calculations.}
\label{fig:invariance}
\end{figure}

In general we expect that the probability $P_a$ will be different when the Hamiltonian is different (e.g., transformed by some symmetry operations). 
However, the acceptor population $P_a$, is invariant under the simultaneous sign change of the excitation energy difference $\Delta$ and the vibrational frequency $\nu$. 
Consider the time traces of the transfer probability $P_a(t)$, Eq.~\eqref{eq:Pa}, shown in Fig.~\ref{fig:invariance} as an example. It is evident from these plots that for the off-resonant transition (i.e., $\nu^2\neq \Delta^2+J^2$), the empty circles for which $\{\Delta/2\pi,\nu/2\pi\}=\{1.2,-0.9\}$kHz give the same probability as the filled circles, for which $\{\Delta/2\pi,\nu/2\pi\}=\{-1.2,0.9\}$kHz, i.e., with opposite signs of both $\Delta$ and $\nu$.
This is also true for the resonant transitions $\nu^2\neq \Delta^2+J^2$ both in the absence of classical noise, indicated by the red empty and filled squares with $\{\Delta/2\pi,\nu/2\pi\}=\{1.2,-0.9\}$kHz and $\{-1.2,0.9\}$kHz, respectively, in Fig.~\ref{fig:invariance}, and in the presence of the classical noise source ($\delta\neq0$, not shown here). 
To understand this invariance, we performed a symmetry-based analysis for both the situation in the absence of noise and the situation in the presence of noise.
As detailed in Appendix~\ref{app:symmetryAnalysis}, we find that if the initial state is an eigenstate of a parity operator and of the Pauli operator $\sigma_z$, and if all coefficients in the initial state have the same phase modulo $\pi$, then the probability $P_a (t)$ is invariant when simultaneously changing the sign of $\Delta$ and $\nu$. This invariance is independent both of the value of the temperature parameter $n_b$ and of whether or not the resonance condition ($\Delta^2+J^2=\nu^2$) is satisfied. This surprising independence derives fundamentally from the inherited parity symmetry and the time-reversal symmetry of the Hamiltonian that fully governs dynamical evolution under a certain initial state associated with an additional symmetry. See Appendix~\ref{app:symmetryAnalysis} for full details.
We finally mention that, for more complex photosynthetic systems or trapped-ion platforms that go beyond the Hamiltonian under our consideration, one could perform similar analysis as to that in Appendix~\ref{app:symmetryAnalysis} to assess whether an invariant property exists or not.

\section{Features of the noisy VAET}
\label{sec:features}

\begin{figure}
\centering
  \includegraphics[width=.9\columnwidth]{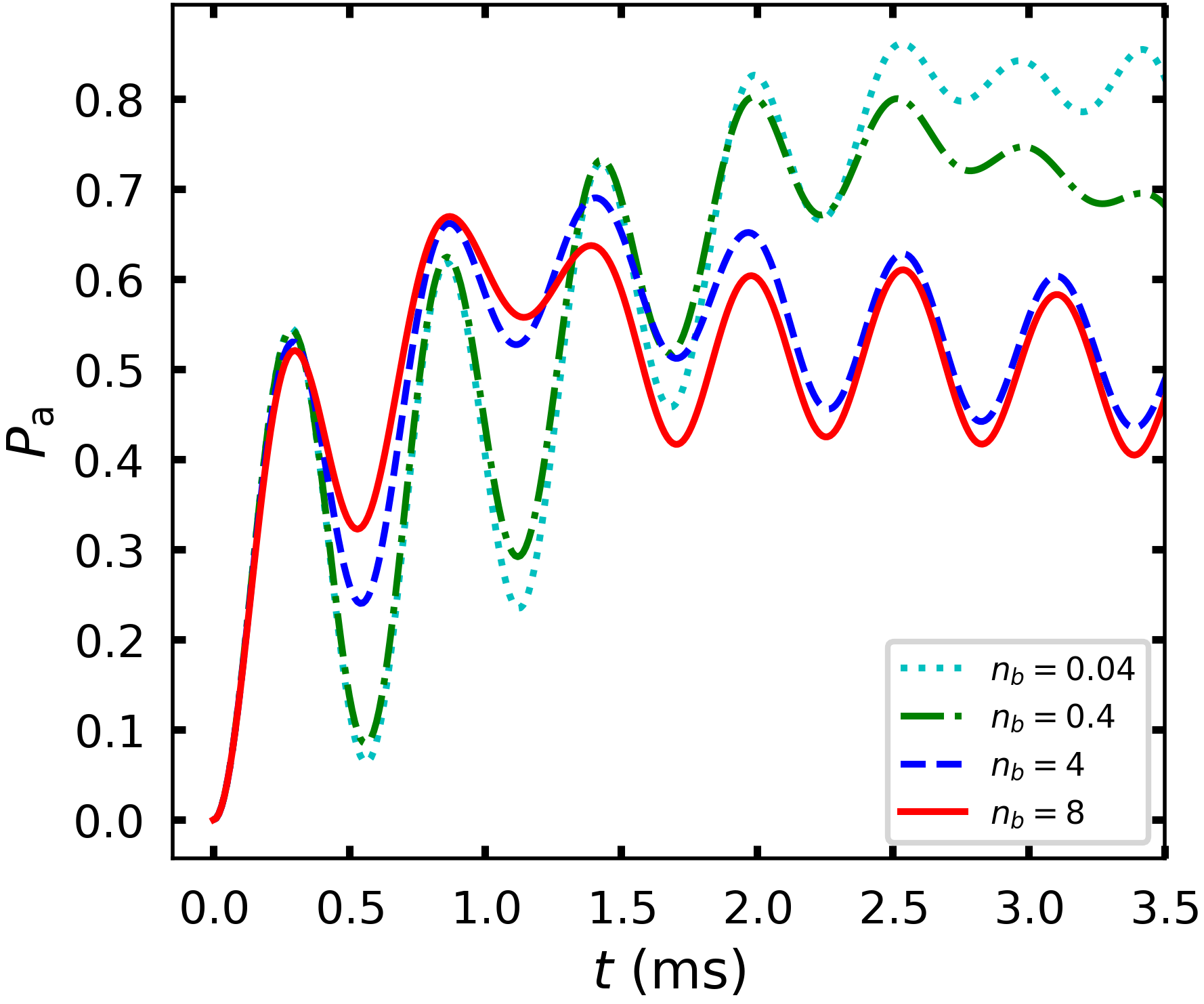} 
\caption{(color online) A sample time evolution of transfer probability between electronic excited states at sites of the donor-acceptor dimer for various values of temperature ($n_b=0.04,0.4,4,8$) in the absence of classical noise ($\delta=0$). The other parameters are $\Delta/2\pi=1.2$kHz, $\nu/2\pi=1.8$kHz, $J/2\pi=1.3$kHz, $\kappa/2\pi=0.229$kHz, and $N_r=800$.}
\label{fig:evolution}
\end{figure}

Here we demonstrate how the basic features of excitonic energy transport are influenced by the interplay between quantum and classical noise present in the dimeric chromophore donor-acceptor system of Eq.~\eqref{eq:Hamiltonian}, considering in particular the effect of the classical stochastic noise on the VAET induced by the quantum noise. 
In Fig.~\ref{fig:evolution} we first present a typical example of the time evolution of the transfer probability in the absence of classical noise. This shows oscillations characterized by the transition frequency $\nu\sim \sqrt{\Delta^2+J^2}$, with a corresponding oscillatory period $2\pi/\nu\sim0.556$ms (approximately four cycles in each period of $2.4$ms). 
Because of the coherent coupling of the donor-acceptor dimer to the underdamped vibration, we expect additional slow oscillations at a frequency $\sim\frac{\kappa J}{\nu}\sqrt{n}$ (period $\sim 6.4$ms).  These occur on a longer time scale than that shown here. 
It is also shown in Fig.~\ref{fig:evolution} that with the increase of the average phonon number $n_b$, that quantifies the temperature of the vibration mentioned in Sec.~\ref{sec:model}, $P_a$ first increases as expected but becomes suppressed at later times. This suppression results from a reduced period of the above-mentioned slow oscillations for an increased $n_b$ which has a relatively higher probability (i.e., $n_b^n/(n_b+1)^{n+1}$) for a larger $n$ state $|n\rangle\langle n|$.
Starting from this reference behavior with no classical noise ($\delta=0$), in the following we shall develop an understanding of the effect of finite classical noise by sweeping the values of both the vibrational frequency ($\nu$) and the donor-acceptor excitation energy difference ($\Delta$).

\begin{figure}
\centering
  \includegraphics[width=.999\columnwidth]{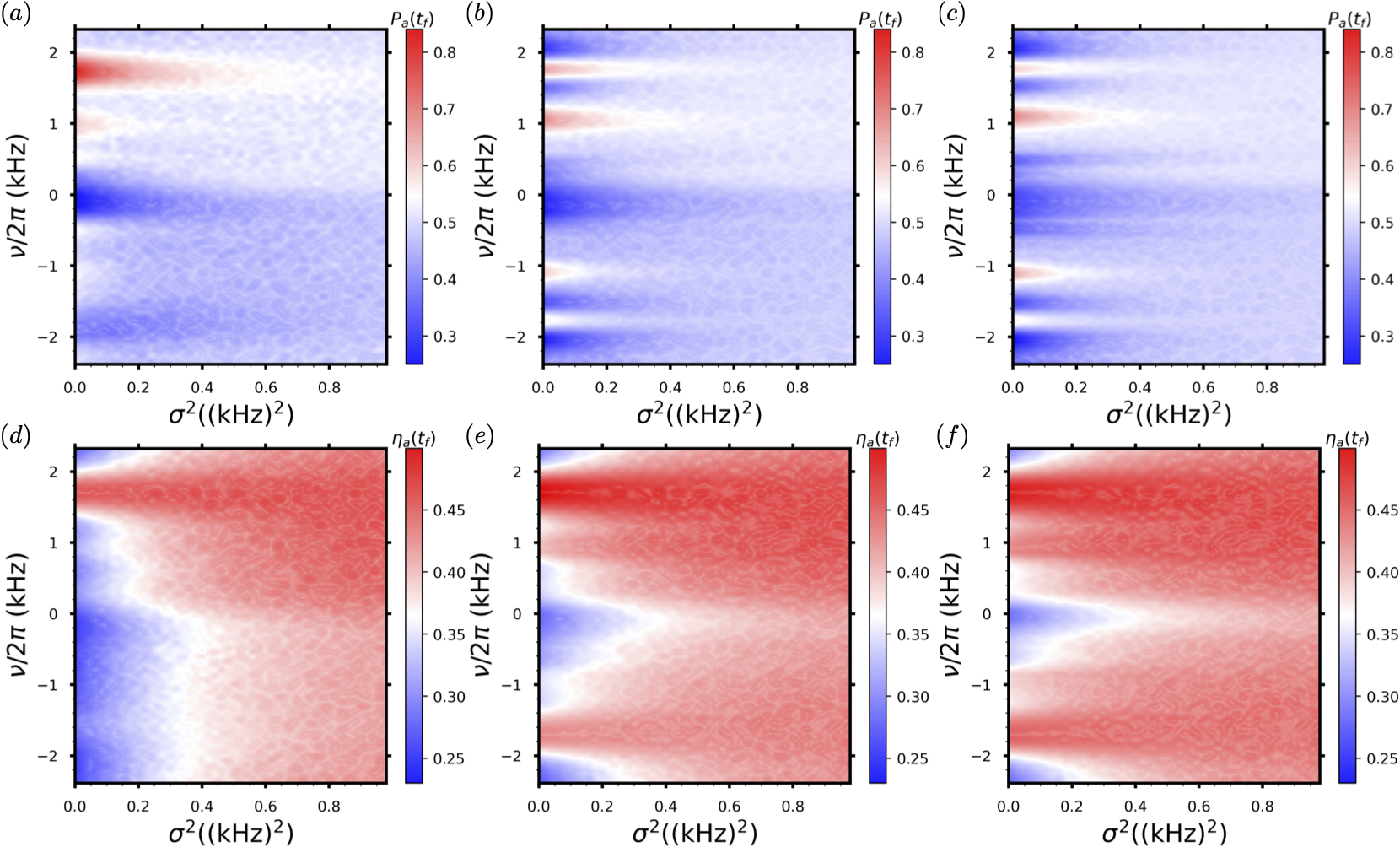} 
\caption{(color online) Probability $P_a(t_f)$ (upper panel a-c) and efficiency $\eta_a(t_f)$ (lower panel d-f) of energy transfer as a function of vibrational mode frequency ($\nu$) and classical noise variance ($\sigma^2$). The detuning between donor and acceptor is $\Delta/2\pi=1.2$kHz. The three upper/lower plots are at different temperatures with mean boson mode occupation number ranging from low ($n_b=0.4$), intermediate ($n_b =4$), and high ($n_b =8$). Other parameters are $t_f=2$ms, $J/2\pi=1.3$kHz, $\kappa/2\pi=0.229$kHz, and $N_r=800$.}
\label{fig:sweepingVibrationFreq}
\end{figure}

\subsection{VAET in the presence of weak noise}

The upper panels in Fig.~\ref{fig:sweepingVibrationFreq} show two-dimensional plots of the acceptor population $P_a$ as a function of the vibrational mode frequency $\nu/2\pi$ and the classical noise variance $\sigma^2$ over the time period $ 0 - t_f$, for $t_f=2$ms and three different temperatures (panels a-c).  The lower panels (d-f) show the corresponding efficiencies $\eta_a$ (accumulated population) over same time period. These plots correspond to 
the weak noise (small variance) regime.
The distinct horizontal bars at resonance, $\nu=\pm\sqrt{\Delta^2+J^2}\sim 2\pi\times1.8$kHz (here $\Delta/2\pi=1.2$kHz, $J/2\pi=1.3$kHz), for example, those in panels (a) and (d), are signatures of VAET corresponding to an energy transfer process assisted by one phonon from the underdamped vibration. 
In addition to these resonant points that show high transfer probability $P_a$ and correspondingly high efficiency $\eta_a$, we also observe VAET processes involving more than one phonon. Specifically, the lower intensity horizontal bars at $\nu\sim\pm2\pi\times 0.9$kHz or  $\nu\sim\pm2\pi\times 0.6$kHz indicate the two- or three-phonon absorption processes that can assist excitonic transfer in the donor-acceptor system.  
The plots show that classical noise injected as a random Gaussian modulation of the energy gap between excitations at the donor and acceptor sites begins to play a role as $\sigma^2$ is increased from zero. As expected, this noise is seen to gradually reduce the extent of VAET as $\sigma^2$ increases.  However the extent of this reduction depends on whether the vibrational frequency is positive or negative, as demonstrated in panels (a) and (b). 
This asymmetry in the degree of vibrational assistance for $\pm\sqrt{\Delta^2+J^2}$ becomes less pronounced as temperature increases (see panels (b, e) ($n_b=4$) and panels (c, f) ($n_b=8$)). 
It is also evident that the effect of VAET is increasingly suppressed as the classical noise variance increases, with both $P_a$ and $\eta_a$ becoming increasingly uniform as a function of the vibrational frequency $\nu$.
The meaning of the negative vibrational frequency ($\nu<0$) that appears in Fig.~\ref{fig:sweepingVibrationFreq} is already mentioned in Sec.~\ref{sec:invariance} above and will be further explained in Appendix~\ref{app:symmetryAnalysis}.

\begin{figure}
\centering
  \includegraphics[width=.999\columnwidth]{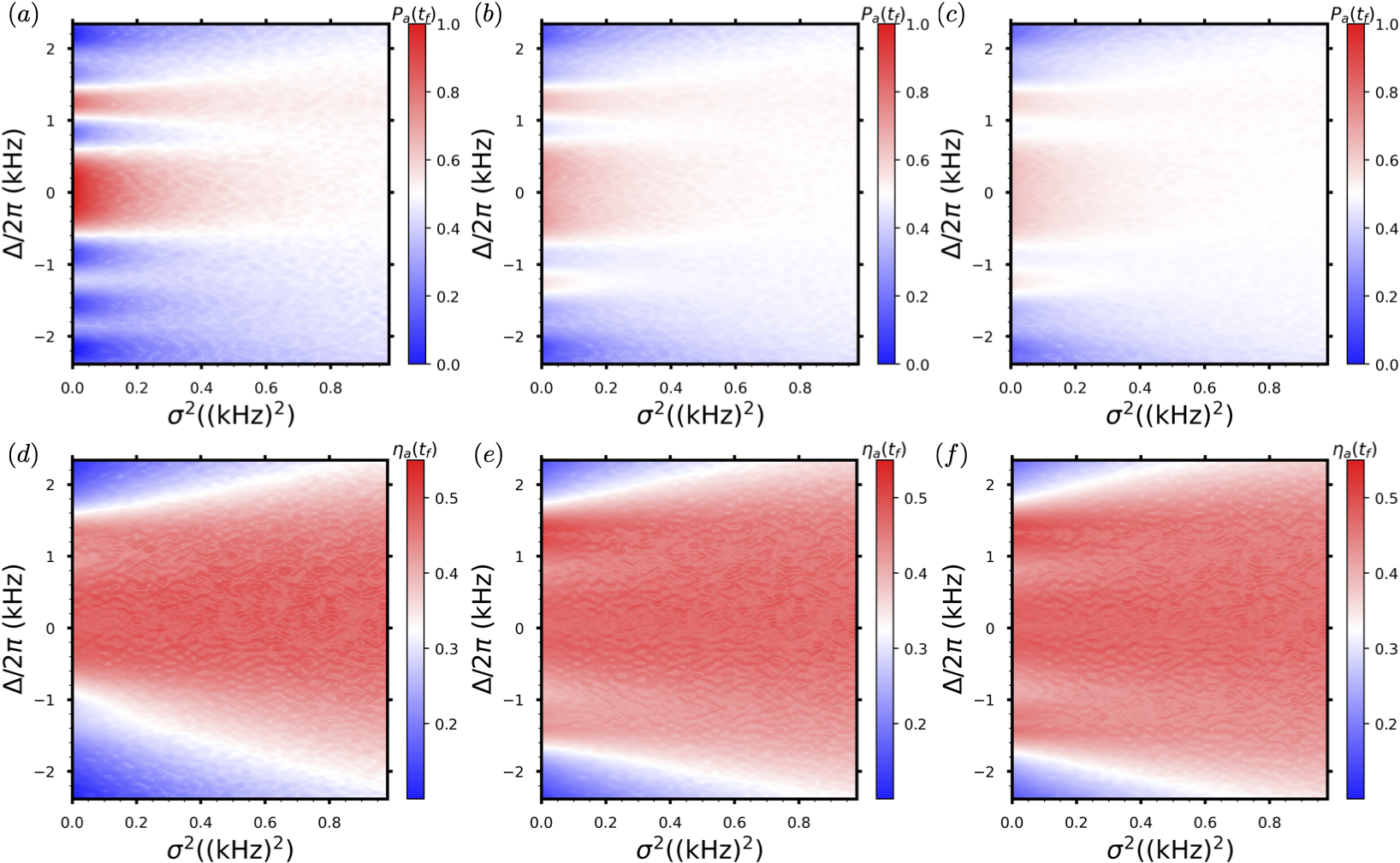} 
\caption{(color online) Probability $P_a(t_f)$ (upper panel a-c) and efficiency $\eta_a(t_f)$ (lower panel d-f) of energy transfer as a function of donor-acceptor energy difference  ($\Delta$) and classical noise variance ($\sigma^2$), for three different temperatures. The three upper/lower plots have mean boson mode occupation number ranging from low ($n_b=0.4$), intermediate ($n_b =4$), and high ($n_b=8$), corresponding to values of temperature $T=0.069\mu$K, $0.387\mu$K, and $1.77\mu$K, respectively. Here the vibrational mode frequency is $\nu/2\pi=1.8$kHz and all other parameters are the same as in Fig~\ref{fig:sweepingVibrationFreq}. Note that a trapped-ion quantum simulator can be operated in the regime where all Hamiltonian parameters are of order a few kHz~\cite{GormanSarovarHaeffner18prx} that leads to low temperatures evaluated above.}
\label{fig:sweepingEnergyDiff}
\end{figure}

In addition to sweeping the vibrational mode frequency in Fig.~\ref{fig:sweepingVibrationFreq}, we have also considered sweeping the donor-acceptor energy difference, $\Delta$, since this may be easier to realize in experiments. 
Fig.~\ref{fig:sweepingEnergyDiff} shows the transfer probability (upper panel a-c) and efficiency (lower panel d-f) as a two-dimensional function now of $\Delta /2\pi$ and of the classical noise variance $\sigma^2$. 
In this case, the optimal transfer efficiency is found at $\Delta=0$ since the donor and acceptor are in resonance here. 
There are still distinct peaks around $\Delta=\pm\sqrt{\nu^2-J^2}\sim\pm2\pi\times 1.2$kHz (with $\nu/2\pi = 1.8$kHz and $J/2\pi=1.3$kHz), see for example panel (a).  
Note that a peak at a positive value of $\Delta$ represents downhill energy transfer with emission of a phonon, so the signatures of these features are more intense. Peaks at negative values of $\Delta$ signify uphill VAET processes assisted by absorption of a phonon from the vibration. 
There are additional peaks observable near $\Delta/2\pi\sim\pm1.8$kHz in Fig.~\ref{fig:sweepingEnergyDiff}(a), implying off-resonant transitions assisted by the vibrational mode $\nu/2\pi=1.8$kHz.
We see again that increasing the variance of the classical noise decreases the VAET signatures [see Fig.~\ref{fig:sweepingEnergyDiff}(b) or (c)], similar to what was observed in Fig.~\ref{fig:sweepingVibrationFreq}. 

Comparing Figs.~\ref{fig:sweepingVibrationFreq}  and \ref{fig:sweepingEnergyDiff}  we see a clear difference in the behavior of the energy transfer efficiency depending on which parameter is swept, i.e., $\nu$ or $\Delta$.  
This difference derives fundamentally from the different aspects of the basic relation $n\nu=\pm\sqrt{\Delta^2+J^2}$ (with $n$ the number of phonons in the vibration), revealed by either fixing the donor-acceptor gap $\Delta$ ($>0$ in Fig.~\ref{fig:sweepingVibrationFreq}) and sweeping $\nu$, or fixing the vibrational frequency $\nu$ ($>0$ in Fig.~\ref{fig:sweepingEnergyDiff}) while sweeping $\Delta$. Sweeping $\nu$  allows the $n>1$, multiple phonons resonances to be seen, while sweeping $\Delta$ allows the $\Delta=0$, donor/acceptor resonance to be seen.

\subsection{Optimal value of the classical noise variance for energy transport efficiency}

\begin{figure}
\centering
  \includegraphics[width=.99\columnwidth]{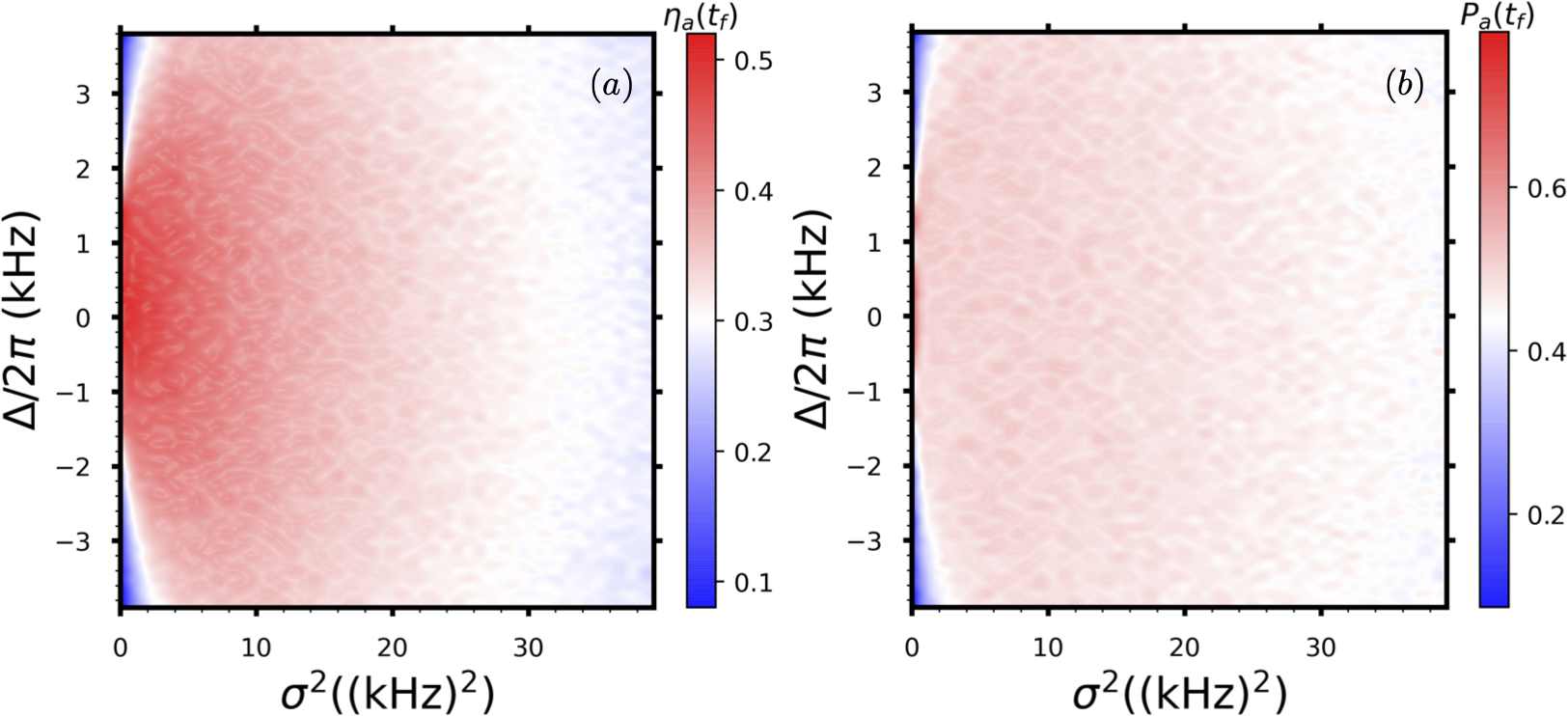} 
\caption{(color online) The turnover of efficiency (a) and probability (b) of the excitation energy transfer as a function of the classical noise variance $\sigma^2$ for larger values of this, with $n_b=4$. All other parameters are the same as in Fig.~\ref{fig:sweepingEnergyDiff}.}
\label{fig:ENAQT}
\end{figure}

The previous plots have focused on the low noise regime. In this regime, the existence of an optimal classical noise variance $\sigma^2$ at which the efficiency is maximized, the key prediction of purely dephasing-enhanced energy transfer~\cite{RebentrostAspuru09njp}, is not evident. Instead, the efficiency for a finite donor-acceptor gap ($\Delta\neq0$) is seen to increase with increasing $\sigma^2$. In order to see a turnover of efficiency with increasing noise variance, one has to study larger values of $\sigma^2$. 
Fig.~\ref{fig:ENAQT}(a) shows plots of the energy transfer efficiency at larger values of the classical noise variance $\sigma^2$.
For a given finite energy detuning between donor and acceptor ($\Delta=\sqrt{\nu^2-J^2}\sim 2\pi\times1.2$kHz), the efficiency is seen to be low for small noise variance $\sigma^2$, to rise as $\sigma^2$ increases, and is now seen to subsequently decrease again at the significantly higher values of $\sigma^2$ used here. 
This behavior is characteristic of the turnover of quantum random walks under dephasing~\cite{Kendon_2007} and has been termed an ENAQT turnover~\cite{RebentrostAspuru09njp}. 
Fig.~\ref{fig:ENAQT}(b) shows the corresponding probability $P_a(t_f)$, which also shows an optimal $\sigma$ value but with a weaker maximum.
It is evident from these plots that the appearance of a quantum Zeno regime at large $\sigma^2$, where the dephasing severely inhibits any amount of energy transfer, is more easily observable for large detuning values $|\Delta|>J$.

We conclude that in the high noise regime, ENAQT dominates and VAET plays little role in energy transfer dynamics. This is expected since vibrational assistance is a resonant phenomenon, and noise broadening of the energy difference between donor and acceptor eliminates a well-defined energy gap for the vibrational mode to be on resonance with.

\section{Discussions and conclusions}
\label{sec:concl}
We have studied the interplay between vibrationally assisted excitonic energy transfer due to coupling of electronically excited states with underdamped vibrational modes and the effect of classical dephasing that has been previously identified as enabling 
energy transfer in certain regimes. We illustrated the interplay between these two processes by considering a dimeric chromophore donor-acceptor dimer system, which provides the basic features of a larger light-harvesting complex that are relevant to the interaction of excitonic and vibrational degrees of freedom.
We find that while addition of the classical noise weakens the VAET processes and destroys the quantum signature of this in the weak noise regime, it can nevertheless enhance the VAET when the noise becomes higher. We also established an invariance property of this dimeric VAET system and provided a symmetry-based  explanation for this.
We look forward to experiments, e.g., on trapped ion emulation platforms, that will demonstrate and validate the results shown here, in particular the interplay between VAET and dephasing-induced optimal values of classical noise, as well as the invariance property.
We expect that the basic features of the interplay between quantum and classical noise, representing coupling with underdamped and overdamped oscillators, the latter confined to dephasing interactions, that have been demonstrated and analyzed here for a typical dimeric chromophore donor-acceptor dimeric system can be manifested also in larger and more complex photosynthetic systems. 
Generalization to other kinds of noise, or to noise injected as a modulation of the vibrational mode frequency or of the exciton-vibration coupling is straightforward. 
We expect that our observation that weak classical noise tends to weaken VAET will generally hold for other kinds of generic noise as well. It would be interesting if this is not true for some highly structured noise with tailored spectrum.

We emphasize that the detrimental effect of the classical noise on the VAET reported in the present work is dependent on frequency of molecular vibration and difference of excitonic energies~\cite{IshizakiFleming21jpcb}. 
Specifically, in Fig.~\ref{fig:sweepingVibrationFreq}(a-c) with a fixed donor-acceptor gap, the detrimental effect of the classical noise manifests itself in different ways for a resonant absorption of either single or multiple phonons from the vibration. 
For the off-resonant cases in Fig.~\ref{fig:sweepingVibrationFreq}(a-c), the classical noise demonstrates a beneficial effect (this is because noise can bring the excitation levels to resonance and thereby enhance the energy transfer). 
In Fig.~\ref{fig:sweepingEnergyDiff}(a-c) with a fixed vibrational mode frequency, the detrimental effect of the classical noise truly depends on the donor-acceptor gap, e.g., downhill $(\Delta>0)$ or uphill $(\Delta<0)$, and resonant or off-resonant transfer. In addition, our observation in Fig.~\ref{fig:sweepingEnergyDiff}(a-c) is consistent with that in Ref.~\onlinecite{IshizakiFleming21jpcb}, namely the vibration plays a minor role in the region of small absorption energy difference but contributes strongly to assist the transfer in the region of large absorption energy. 

We also comment on the relationship between our work and Ref.~\onlinecite{FujihashiFlemingIshizaki15jcp}, which studied a similar problem of excitation energy transfer of a dimer under the influence of both a vibration and an environment. 
In that work, it is claimed that the transfer dynamics are dominated by the environment and the vibrational modes contribute to acceleration the energy transfer only slightly. While this conclusion is consistent with the conclusions drawn in the present work, 
we would like to point out a few differences between Ref. \onlinecite{FujihashiFlemingIshizaki15jcp} and our work.
The first one is the theoretical methods employed. The environment and the vibration were taken into account in Ref.~\onlinecite{FujihashiFlemingIshizaki15jcp} via relaxation functions in the form of either an exponential decay or Brownian oscillator model, respectively, which is different from the Gaussian white noise or the quantized oscillator included in a straightforward way into the Hamiltonian [i.e., Eq.~(\ref{eq:Hamiltonian})] in our work. 
The second difference is that  Ref.~\onlinecite{FujihashiFlemingIshizaki15jcp} considered a specific case, namely, resonant downhill energy transfer affected by a vibration and an environment.
In contrast, we consider the effect of finite classical noise in a more complete way by sweeping the values of the vibrational frequency, the donor-acceptor excitation energy difference, or the variance of the classical noise, as shown explicitly in the weak-noise regime in Figs.~\ref{fig:sweepingVibrationFreq} and \ref{fig:sweepingEnergyDiff}.

Finally, we note that in addition to the trapped-ion platform that we focus on in our work, the circuit-QED platform could also be used for the implementation of our predictions. The effects of classical noise on energy transfer have already been studied on this platform ~\cite{Potocnik_2018}, where individual superconducting qubits played the role of energy sites. As in that case, the interaction between sites can be achieved by cavity-mediated, or direct, coupling between qubits. The new element, the vibrational mode, can be modeled by a mode of a transmission line resonator. A key difference in superconducting platforms, is that the usual coupling between qubits and resonator modes is of the form $\sigma^+a + \sigma_-a^{\dagger}$, which is different from the usual VAET coupling considered here, $\sigma_z (a+a^{\dagger})$. However, this coupling could be engineered by operating in the dispersive regime \cite{Blais_2004} and linearizing the interaction around a large classical pump. 

\acknowledgements

We thank Robert Cook, Joseph Broz, and Hartmut H\"{a}ffner for helpful discussions. 
Work at the University of California, Berkeley was supported by the U.S. Department of Energy (DOE), Office of Science, Basic Energy Sciences (BES) under Award DE-SC0019376. 
Work at Sandia National Laboratories was supported by the U.S. Department of Energy, Office of Science, Basic Energy Sciences, Chemical Sciences, Geosciences, and Biosciences Division. 
Sandia National Laboratories is a multimission laboratory managed and operated by NTESS, LLC., a wholly owned subsidiary of Honeywell International, Inc., for the U.S. DOE's NNSA under contract DE-NA-0003525. This paper describes objective technical results and analysis. Any subjective views or opinions that might be expressed in the paper do not necessarily represent the views of the U.S. Department of Energy or the United States Government.

\appendix

\begin{figure}
\centering
  \includegraphics[width=.9\columnwidth]{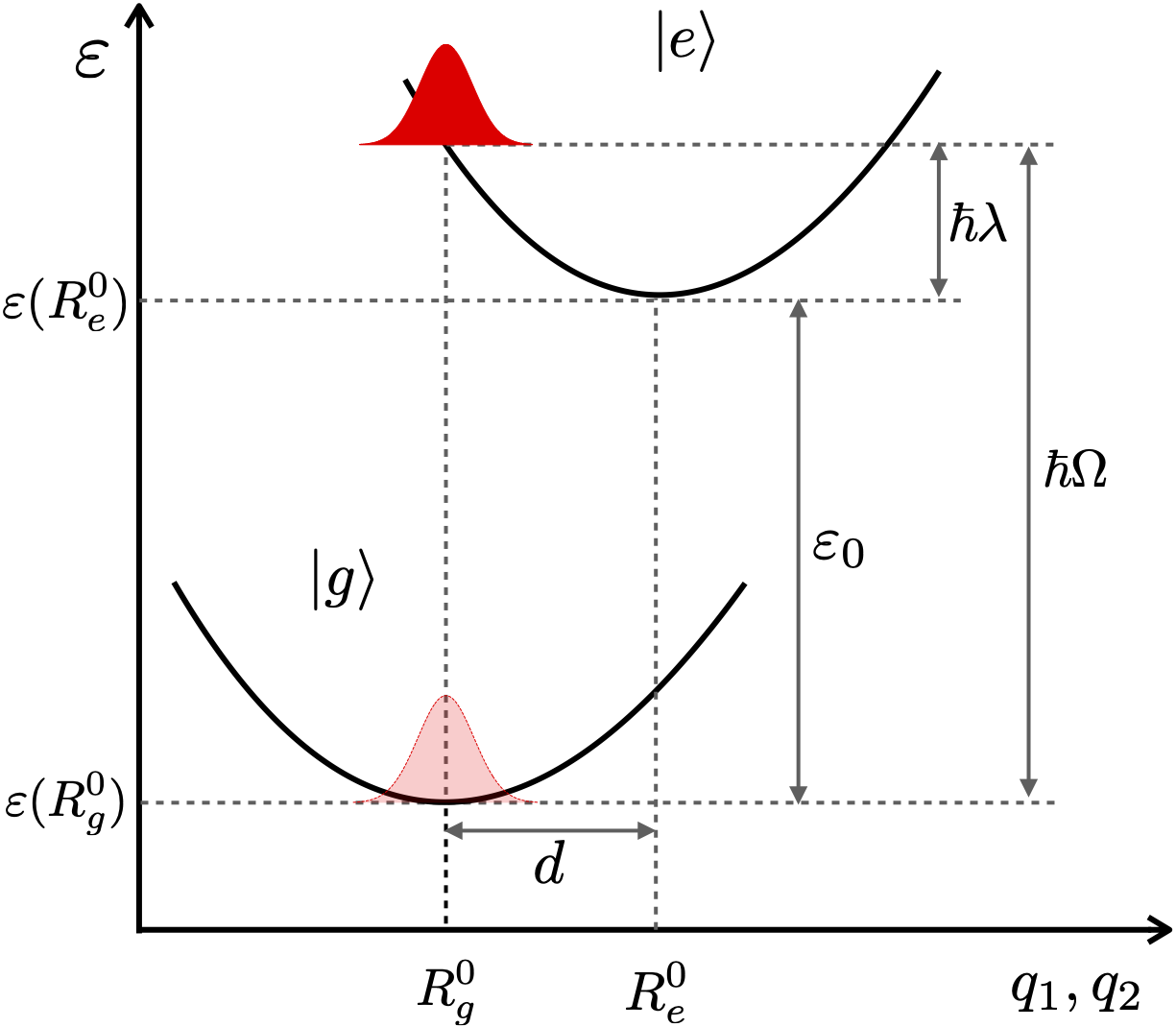} 
\caption{Schematic of the electronic excitation and reorganization processes with the Franck-Condon transition energy $\hbar\Omega$ and the reorganization energy $\hbar\lambda$.}
\label{fig:electro_nuclear}
\end{figure}

\section{Derivation of the electron-vibration interaction\label{app:technique}}

In this appendix, we demonstrate how to connect the Hamiltonian for light harvesting systems that is commonly studied for the natural systems with an equivalent description derived for emulation of these by artificial systems such as trapped ions. 

\subsection{Equivalent descriptions with different coupling forms}

To demonstrate in a simple manner an equivalence between two Hamiltonian descriptions that consider a coupling of a vibration to either an excited state or both ground and excited states~\cite{IshizakiFleming10}, respectively, we take the molecular system shown in Fig.~\ref{fig:electro_nuclear} as an example. This is described by the Hamiltonian
\begin{eqnarray}
H(q_1,q_2) &=& (\frac{p^2}{2m} +\frac{m\omega^2}{2} q_1^2) |g\rangle\langle g| \notag\\
&&+ (\frac{p^2}{2m} +\frac{m\omega^2}{2} q_2^2 + \varepsilon_0) |e\rangle\langle e| .
\end{eqnarray}
Here the zero of energy is defined to be at the minimum of the nuclear potential energy surface for the electronic ground state $|g\rangle$. 
According to Fig.~\ref{fig:electro_nuclear}, we define the Franck-Condon transition energy $\hbar \Omega$ and reorganization energy $\hbar \lambda$ as
\begin{eqnarray}
\hbar\Omega &=& \varepsilon_0 + m\omega^2 q_1 d, \\
\hbar\lambda &=&  \frac{1}{2} m\omega^2 d^2,
\end{eqnarray}
respectively. 

Written in terms of the center of mass and difference coordinates $\bar{q}$ and $d$, defined as $\bar{q}=\frac{q_2+q_1}{2}$ and $d=\frac{q_2-q_1}{2}$ with $q_1$ and $q_2$ the potential surface coordinate of the ground and excited states, respectively, the above Hamiltonian becomes
\begin{eqnarray}
H(\bar{q},d) &=& [\frac{p^2}{2m} +\frac{m\omega^2}{2} (\bar{q}-d)^2 ] |g\rangle\langle g| \notag\\
&&+ [\frac{p^2}{2m} +\frac{m\omega^2}{2} (\bar{q}+d)^2 + \varepsilon_0 ] |e\rangle\langle e| \notag\\
&=&
(\frac{p^2}{2m} +\frac{m\omega^2}{2}\bar{q}^2) (|g\rangle\langle g| + |e\rangle\langle e|) 
 + \frac{m\omega^2}{2}  (d^2 \notag\\
&&-2\bar{q}d) |g\rangle\langle g| 
+ [\frac{m\omega^2}{2} (d^2+2\bar{q}d) + \varepsilon_0] |e\rangle\langle e| \notag\\
&=&
\frac{p^2}{2m} +\frac{m\omega^2}{2}\bar{q}^2 + \frac{m\omega^2}{2} d^2 +\varepsilon_0 |e\rangle\langle e| 
+ m\omega^2\bar{q}d \sigma_z .  \notag\\
\label{eq:couplingGE}
\end{eqnarray}
Here $\sigma_z=|e\rangle\langle e| - |g\rangle\langle g|$. It is then obvious that both the ground and the first excited states are coupled to the vibration. 
An alternative equivalent form $H(q_1,d)$ is obtained by substituting $q_2=q_1+2d$ into $H(q_1,q_2)$. This yields
\begin{eqnarray}
H(q_1,d) &=& (\frac{p^2}{2m} +\frac{m\omega^2}{2} q_1^2) |g\rangle\langle g|  \notag\\
&&+ [\frac{p^2}{2m} +\frac{m\omega^2}{2} (q_1+2d)^2 + \varepsilon_0] |e\rangle\langle e| \notag\\
&=&
(\frac{p^2}{2m} +\frac{m\omega^2}{2} q_1^2) (|g\rangle\langle g| + |e\rangle\langle e|)
+ \varepsilon_0 |e\rangle\langle e| \notag\\
&&+ 2m\omega^2q_1d |e\rangle\langle e| + 2m\omega^2d^2|e\rangle\langle e| \notag\\
&=&
\frac{p^2}{2m} +\frac{m\omega^2}{2} q_1^2 + \varepsilon_0 |e\rangle\langle e| 
+ 2m\omega^2q_1d |e\rangle\langle e| \notag\\
&&+ 2m\omega^2d^2 |e\rangle\langle e| .
\label{eq:couplingE}
\end{eqnarray}
It is evident that in this form the vibration is coupled only to the excited state. 

As demonstrated, $H(\bar{q},d)$ and $H(q_1,d)$ are both equivalent to $H(q_1,q_2)$, despite containing different forms for the coupling between the vibration and the electron. 
This equivalence can be also illustrated by further substituting $q_1=\bar{q}-d$,
\begin{eqnarray}
H(q_1,d) &=& \frac{p^2}{2m} +\frac{m\omega^2}{2} (\bar{q}-d)^2 + \varepsilon_0 |e\rangle\langle e| \notag\\
&&+ 2m\omega^2 (\bar{q}-d) d |e\rangle\langle e| + 2m\omega^2d^2 |e\rangle\langle e| \notag\\
&=&
 \frac{p^2}{2m} +\frac{m\omega^2}{2} \bar{q}^2 + \frac{m\omega^2}{2} d^2 - m\omega^2\bar{q}d 
+ \varepsilon_0 |e\rangle\langle e|  \notag\\
&& +  2m\omega^2 (\bar{q} d - d^2)  |e\rangle\langle e| + 2m\omega^2d^2|e\rangle\langle e| \notag\\
&=&
\frac{p^2}{2m} +\frac{m\omega^2}{2} \bar{q}^2 + \frac{m\omega^2}{2} d^2 + \varepsilon_0 |e\rangle\langle e| \notag\\
&&+ m\omega^2\bar{q}d (|e\rangle\langle e| - |g\rangle\langle g|) 
= H(\bar{q},d) .
\end{eqnarray}

\subsection{Connection to our model}

The lesson that we learn from the above analysis is that two alternative forms of a Hamiltonian with coupling of the vibration to either the ground state or both excited and ground states can be appropriate for describing the system. In particular, one needs to be careful about which additional energetic terms are needed in order to establish an equivalence between the two forms.  

To make the connection of the above-demonstrated equivalence with our current model more apparent, we rewrite Eqs.~(\ref{eq:couplingGE}) and (\ref{eq:couplingE}) by performing a shift of the zero of energy to the average of the minima of the two potential surfaces and further quantizing the two quadratic Hamiltonians. 
When the zero of energy is shifted to $\frac{\varepsilon_0}{2}$, 
Eq.~(\ref{eq:couplingGE}) can be rewritten as
\begin{eqnarray}
H(\bar{q},d) &=&
\frac{p^2}{2m} +\frac{m\omega^2}{2}\bar{q}^2 + \frac{m\omega^2}{2} d^2 \notag\\
&&+\frac{\varepsilon_0}{2} ( |e\rangle\langle e| - |g\rangle\langle g|)
+ m\omega^2\bar{q}d \sigma_z \notag\\
&=&
\frac{p^2}{2m} +\frac{m\omega^2}{2}\bar{q}^2 + \frac{m\omega^2}{2} d^2 
+\frac{\varepsilon_0}{2} \sigma_z
+ m\omega^2\bar{q}d \sigma_z .\notag\\
\label{eq:couplingGE_shift}
\end{eqnarray}
A further quantization of the harmonic oscillator, i.e., $\frac{p^2}{2m} +\frac{m\omega^2}{2}\bar{q}^2\rightarrow\hbar a^{\dagger}a$ and $\bar{q}\rightarrow a^{\dagger}+a$, yields the Hamiltonian
\begin{eqnarray}
H &=& \hbar\omega a^{\dagger}a + \frac{\varepsilon_0}{2}\sigma_z -\kappa (a^{\dagger}+a) \sigma_z + \frac{m\omega^2d^2}{2} \notag\\
&=& \hbar\omega a^{\dagger}a + \frac{\varepsilon_0}{2}\sigma_z -\kappa (a^{\dagger}+a) \sigma_z + \frac{\kappa^2}{2m\omega^2} ,
\label{eq:couplingGE_shift_quanti}
\end{eqnarray}
where $\kappa=-m\omega^2d$. 
Similarly, Eq.~(\ref{eq:couplingE}) under the consideration of the same shifted zero of energy becomes
\begin{eqnarray}
H(q_1,d) &=& 
\frac{p^2}{2m} +\frac{m\omega^2}{2} q_1^2 + \frac{\varepsilon_0}{2} (|e\rangle\langle e| - |g\rangle\langle g|) \notag\\
&&+ 2m\omega^2q_1d |e\rangle\langle e| + 2m\omega^2d^2 |e\rangle\langle e| \notag\\
&=& 
\frac{p^2}{2m} +\frac{m\omega^2}{2} q_1^2 + \frac{\varepsilon_0}{2} \sigma_z \notag\\
&&+ 2m\omega^2q_1d |e\rangle\langle e| + 2m\omega^2d^2 |e\rangle\langle e|.
\label{eq:couplingE_shift}
\end{eqnarray}
Quantization of Eq.~\eqref{eq:couplingE_shift} using $\frac{p^2}{2m} +\frac{m\omega^2}{2} q_1^2\rightarrow\hbar \tilde{a}^{\dagger}\tilde{a}$ and $q_1\rightarrow\tilde{a}^{\dagger}+\tilde{a}$ leads to
\begin{eqnarray}
\tilde{H} &=&
 \hbar \tilde{a}^{\dagger}\tilde{a} + \frac{\varepsilon_0}{2} \sigma_z - 2\kappa (\tilde{a}^{\dagger}+\tilde{a}) |e\rangle\langle e| + \frac{2\kappa^2}{m\omega^2} |e\rangle\langle e|  \notag\\
&=&
 \hbar \tilde{a}^{\dagger}\tilde{a} + (\frac{\varepsilon_0}{2} + \frac{2\kappa^2}{m\omega^2}) |e\rangle\langle e| 
 - 2\kappa (\tilde{a}^{\dagger}+\tilde{a}) |e\rangle\langle e|  \notag\\
&& - \frac{\varepsilon_0}{2} |g\rangle\langle g| .
\label{eq:couplingE_shift_quanti}
\end{eqnarray}
Comparison between Eq.~(\ref{eq:couplingGE_shift_quanti}) and Eq.~(\ref{eq:couplingE_shift_quanti}) clearly shows
the relationship between equivalent descriptions with different forms of the electron-vibration coupling.

Finally we illustrate the connection between the Huang-Rhys parameter in the natural systems and the site-vibration coupling of the simulation systems. The Huang-Rhys parameter $D$ is a dimensionless factor related to the scaled mean square displacement
\begin{eqnarray}
D&=&\frac{d'^2m\omega}{2\hbar} ,
\end{eqnarray}
where $d'=q_2-q_1=2d$. This yields the displacement $d'=\sqrt{\frac{2\hbar D}{m\omega}}$. 
The interaction term between the local excitonic degree of freedom and the vibration in Eq.~(\ref{eq:couplingGE_shift}) has the form 
\begin{eqnarray}
&&m\omega^2 \bar{q} d \sigma_z \notag\\
&=& m\omega^2\times  \frac{1}{2}\sqrt{\frac{2\hbar D}{m\omega}} 
\times \sqrt{\frac{\hbar}{2m\omega}} (a^{\dagger} + a)
\times (|e\rangle\langle e| -|g\rangle\langle g|) \notag\\
&=& \frac{1}{2}\hbar\omega \sqrt{D} (a^{\dagger} + a) \sigma_z .
\end{eqnarray}
Equivalently, if we consider the alternative coupling form in Eq.~(\ref{eq:couplingE_shift}), this interaction is given by
\begin{eqnarray}
&&2m \omega^2 q_1 d  |e\rangle\langle e| \notag\\
&=& m\omega^2 \sqrt{\frac{2\hbar D}{m\omega}} 
\times \sqrt{\frac{\hbar}{2m\omega}} (a^{\dagger} + a) 
\times |e\rangle\langle e| \notag\\
&=& \hbar\omega \sqrt{D} (a^{\dagger} + a) |e\rangle\langle e| .
\end{eqnarray}
This implies the site-vibration coupling parameter $\kappa=-\sqrt{D}\hbar\omega/2$.

\section{Symmetry analysis on the VAET system \label{app:symmetryAnalysis}}

In this appendix we present a symmetry-based analysis to support the invariance property of our VAET system both in the absence and in the presence of the classical noise. 

\subsection{Absence of classical noise}

We first consider the case of the absence of classical noise, $\bar{H}=\tilde{H}(\delta=0)$. 
Given the effective Hamiltonian in Eq.~(\ref{eq:Hamiltonian_SingleEx}) and the initial state $|\phi_0\rangle (=|eg,n\rangle)$, the probability of being in the target state at time $t$ is $P(t)=\langle\phi_0| e^{i \bar{H} t} \Pi_a e^{-i \bar{H} t}|\phi_0\rangle$, where the  $ \Pi_a=|ge\rangle\langle ge|$ is the projection onto the target state. 
The extension to a thermal state of the vibration $\rho_b=\sum_{n=0}^{\infty} \frac{n_b^n}{(n_b+1)^{n+1}} |n\rangle\langle n|$ is straightforward.
We are going to transform the Hamiltonian by conjugating it with different symmetry operations, which corresponds
to changing the signs of some subsets of the four parameters $\Delta$, $J$, $\nu$, and $\kappa$.
We will show that when the initial state $|\phi_0\rangle$ is an eigenstate of the relevant symmetry operations, $P(t)$ is invariant under the corresponding sign changes.

The first symmetry is the parity corresponding to a sign change of the exciton-vibration coupling constant $\kappa$. Under this sign change the Hamiltonian $\bar{H}$ becomes
$H_1=\frac{\Delta}{2}\tilde{\sigma}_z +\frac{J}{2}\tilde{\sigma}_x +\nu a^{\dagger}a -\frac{\kappa}{2}\tilde{\sigma}_z(a+a^{\dagger})$. 
The physical meaning of this sign change is to invert the coordinate of the harmonic oscillator (i.e., a parity operation). If our initial state treats both directions of the coordinate equally, then we expect that inverting the coordinate will have no effect on the system dynamics other than inverting the spatial coordinates.
We define the parity operator $\Upsilon$ just as for the spatial coordinate of the 1D harmonic oscillator, i.e.,
\begin{equation}
\Upsilon|n\rangle= \begin{cases}
-|n\rangle & \quad n \text{ is odd}, \\[1ex]
|n\rangle & \quad n \text{ is even},
\end{cases}
\end{equation}
so the new Hamiltonian can be written as $H_1=\Upsilon^{\dagger} \bar{H} \Upsilon$. The probability of being in the target state then becomes
\begin{eqnarray}
P_1(t)&=&\langle\phi_0| e^{iH_1t} \Pi_a e^{-iH_1t} |\phi_0\rangle \notag\\
&=& \langle\phi_0| e^{i\Upsilon^{\dagger} \bar{H} \Upsilon t} \Pi_a e^{-i\Upsilon^{\dagger} \bar{H}\Upsilon t} |\phi_0\rangle \notag\\
&=& \langle\phi_0| \Upsilon^{\dagger}e^{i \bar{H} t} \Upsilon \Pi_a \Upsilon^{\dagger} e^{-i \bar{H} t}\Upsilon |\phi_0\rangle \notag\\
&=& \langle\phi_0| \Upsilon^{\dagger}e^{i \bar{H} t} \Pi_a e^{-i \bar{H} t}\Upsilon |\phi_0\rangle , 
\end{eqnarray}
where we have used the identity $e^{i\Upsilon^{\dagger} \bar{H}\Upsilon t}=\Upsilon^{\dagger}e^{i \bar{H} t}\Upsilon$ and the fact that $\Upsilon \Pi_a \Upsilon^{\dagger} =\Pi_a$. 
Since the eigenvalue of the parity operator $\Upsilon$ is $\pm1$, $P_1(t)=P(t)$ if $|\phi_0\rangle$ is an eigenstate of $\Upsilon$, i.e., $P(t)$ is invariant under the sign change of $\kappa$.

Similarly, a sign change of $J$ leads to 
$H_2=\frac{\Delta}{2}\tilde{\sigma}_z -\frac{J}{2}\tilde{\sigma}_x +\nu a^{\dagger}a -\frac{\kappa}{2}\tilde{\sigma}_z(a+a^{\dagger})$, 
which corresponds to redefining the ground state with a minus sign, i.e., $-|eg\rangle$ or $-|ge\rangle$ for $\Delta<0$ or $>0$, respectively. 
Unless the initial state has nonzero coefficients on both $|eg\rangle$ and $|ge\rangle$, we expect this redefinition to have no effect on the system dynamics.
The Hamiltonian $H_2$ is obtained by conjugating $\bar{H}$ with $\tilde{\sigma}_z$ (i.e., $H_2=\tilde{\sigma}_z^{\dagger}\bar{H}\tilde{\sigma}_z$). So the target state probability now becomes
\begin{eqnarray}
P_2(t)&=&\langle\phi_0| e^{iH_2t} \Pi_a e^{-iH_2t} |\phi_0\rangle \notag\\
&=& \langle\phi_0| e^{i\tilde{\sigma}_z^{\dagger} \bar{H} \tilde{\sigma}_z t} \Pi_a e^{-i\tilde{\sigma}_z^{\dagger} \bar{H}\tilde{\sigma}_z t} |\phi_0\rangle \notag\\
&=& \langle\phi_0| \tilde{\sigma}_z^{\dagger}e^{i \bar{H} t} \tilde{\sigma}_z \Pi_a \tilde{\sigma}_z^{\dagger} e^{-i \bar{H} t}\tilde{\sigma}_z |\phi_0\rangle \notag\\
&=& \langle\phi_0| \tilde{\sigma}_z^{\dagger}e^{i \bar{H} t} \Pi_a e^{-i \bar{H} t} \tilde{\sigma}_z |\phi_0\rangle .
\end{eqnarray}
Therefore when the initial state $|\phi_0\rangle$ is an eigenstate of $\tilde{\sigma}_z$, $P(t)$ is invariant under the sign change of $J$. 

The last symmetry to be discussed is the time reversal symmetry. 
The physical meaning of the sign change of the entire Hamiltonian $\bar{H}$, namely, $H_3=-\bar{H}$, is most apparent when we look at the time evolution operator $U_3=e^{-i H_3 t} =e^{-i H (-t)}$. Evolving $H_3$ forward in time corresponds to evolving the original Hamiltonian $\bar{H}$ backward in time. Conjugating the forward time evolution operator $e^{-i H t}$ with the time reversal operator $\Theta$ will give us the backward time evolution operator, which is the forward time evolution operator of $H_3$, i.e., 
$\Theta^{\dagger} e^{-i \bar{H} t} \Theta = e^{i \bar{H} t} = e^{-i H_3 t}$.
In the Hilbert space of our VAET system, the effect of $\Theta$ acting on a state is to replace all the coefficients of the basis states by their complex conjugates, i.e.,
$\Theta \sum_{\alpha= eg,ge}\sum_n c_{\alpha,n} |\alpha,n\rangle = \sum_{\alpha= eg,ge} \sum_n c^*_{\alpha,n} |\alpha,n\rangle$.
Note that $\Theta$ is an antiunitary (antilinear and unitary) operator, so one has to be careful when using this with the Dirac bracket notation.
The target state probability under the time-reversed dynamics is
\begin{eqnarray}
P_3(t)&=&\langle\phi_0| e^{i H_3 t} \Pi_a e^{-i H_3 t} |\phi_0\rangle \notag\\
&=& \langle\phi_0| \Theta^{\dagger}e^{i \bar{H} t} \Theta \Pi_a \Theta^{\dagger} e^{-i \bar{H} t} \Theta |\phi_0\rangle \notag\\
&=& \langle\phi_0| \Theta^{\dagger}e^{i \bar{H} t} \Pi_a e^{-i \bar{H} t} \Theta |\phi_0\rangle .
\end{eqnarray}
If all the coefficients in the initial state have the same phase (mod $\pi$), then complex conjugation just adds an overall phase factor, i.e.,
$\Theta |\phi_0\rangle=e^{i\phi}|\phi_0\rangle$ and $\langle\phi_0| \Theta^{\dagger} = e^{-i\phi}|\phi_0\rangle$. Since these two phase factors cancel each other out, we arrive at the result that
$P(t)$ is invariant under sign change of the entire Hamiltonian when 
all coefficients in the initial state $|\phi_0\rangle$ have the same phase modulo $\pi$.

For initial states $|\phi_0\rangle=|eg,n\rangle$ we can combine the results from the above three symmetry analyses, obtaining relevant combinations of the three possible sign changes. 
Furthermore, the above arguments and conclusions also hold for thermal states of the vibration, $\rho_b=\sum_{n=0}^{\infty} \frac{n_b^n}{(n_b+1)^{n+1}} |n\rangle\langle n|$ given that a thermal states is a superposition of many Fock states (i.e., phonon number states $|n\rangle\langle n|$). 
In particular, it is useful to understand why putting a minus sign on the vibrational frequency $\nu$ yields the same result as putting a minus sign on excitonic detuning $\Delta$. 
We can see this by realizing that a simultaneous sign change on $\nu$ and $\Delta$ 
in $\bar{H}=\tilde{H}(\delta=0)=\frac{1}{2}\Delta\tilde{\sigma}_z + \frac{1}{2}J\tilde{\sigma}_x + \nu a^{\dagger}a +  \frac{1}{2}\kappa\tilde{\sigma}_z(a+a^{\dagger})$ 
is equivalent to first changing the sign of the entire Hamiltonian $\bar{H}$ (time-reversal operation), and then reverting the signs of both $\kappa$ (parity operation) and $J$ ($\tilde{\sigma}_z$ operation) back to their original values.

\subsection{Presence of classical noise}

To show that the invariance property also holds in the presence of classical noise, 
we have numerically demonstrated that the same results can be obtained via consideration of the Lindblad master equation.
This demonstration supports the equivalence between the Lindblad equation and the average behavior of adding Gaussian white noise to Hamiltonian~\cite{Ban10pla}. 
We can therefore perform the symmetry analysis on the Lindblad equation
$\dot{\rho}(t)={\cal L}\rho=-i[\bar{H},\rho(t)]+\gamma[\sigma_z,[\sigma_z,\rho(t)]]$, where $\gamma$ is proportional to the variance of the Gaussian white noise, and the probability of being in the target state is given by $P(t) = {\rm Tr} (\Pi_a e^{-i{\cal L} t } \rho_0)$.

For the parity symmetry that changes the Hamiltonian to $\Upsilon^{\dagger} \bar{H} \Upsilon$, we have
\begin{eqnarray}
{\cal L}_1\rho &=& -i[\Upsilon^{\dagger} \bar{H} \Upsilon,\rho(t)]+\gamma,\sigma_z,[\sigma_z,\rho(t)]] \notag\\
&=& -i \Upsilon^{\dagger} [ \bar{H} , \Upsilon \rho(t) \Upsilon^{\dagger}] \Upsilon 
-\gamma \Upsilon^{\dagger} [\sigma_z,[\sigma_z, \Upsilon \rho(t) \Upsilon^{\dagger} ]]\Upsilon \notag\\
&=& \Upsilon^{\dagger} {\cal L} \Upsilon \rho .
\end{eqnarray}
This gives rise to the time evolution
\begin{eqnarray}
e^{-i{\cal L}_1 t} &=& \sum_{n=0}^{\infty} \frac{1}{n!} (-i {\cal L}_1 t)^n 
=  \sum_{n=0}^{\infty} \frac{1}{n!} (-i \Upsilon^{\dagger} {\cal L} \Upsilon t)^n \notag\\
&=& \Upsilon^{\dagger} e^{-i{\cal L} t} \Upsilon .
\end{eqnarray}
Therefore $P'_1(t) = {\rm Tr} (\Pi_a e^{-i{\cal L}_1 t } \rho_0)
={\rm Tr} (\Pi_a  e^{-i{\cal L} t } \Upsilon \rho_0 \Upsilon^{\dagger})$. 
We then arrive at an analogous conclusion to 
the noiseless case, namely that
$P(t)$ is invariant under sign change of $\kappa$ when $\Upsilon \rho(0) \Upsilon^{\dagger} = \rho_0$.

Similarly, for the symmetry that leads to the transformed Hamiltonian $\tilde{\sigma}_z^{\dagger}\bar{H}\tilde{\sigma}_z$, 
\begin{eqnarray}
{\cal L}_2\rho &=& -i[\sigma_z^{\dagger} \bar{H} \sigma_z,\rho(t)]+\gamma,\sigma_z,[\sigma_z,\rho(t)]] \notag\\
&=& -i \sigma_z^{\dagger} [ \bar{H} , \sigma_z \rho(t) \sigma_z^{\dagger}] \sigma_z 
-\gamma \sigma_z^{\dagger} [\sigma_z,[\sigma_z, \sigma_z \rho(t) \sigma_z^{\dagger} ]]\sigma_z \notag\\
&=& \sigma_z^{\dagger} {\cal L} \sigma_z \rho ,
\end{eqnarray}
which leads to
$P'_2(t)= {\rm Tr} (\Pi_a e^{-i{\cal L}_2 t } \rho_0) ={\rm Tr} (\Pi_a e^{-i{\cal L} t } \sigma_z \rho_0 \sigma_z^{\dagger} )$. 
We then conclude that when $\sigma_z \rho_0 \sigma_z^{\dagger} = \rho_0$,  
$P(t)$ is invariant under a sign change of $J$. 

For the time reversal symmetry we have, 
\begin{eqnarray}
{\cal L}_3\rho &=& -i[\Theta^{\dagger} \bar{H} \Theta,\rho(t)]+\gamma,\sigma_z,[\sigma_z,\rho(t)]] \notag\\
&=& -i[-\bar{H} ,\rho(t)]+\gamma,\sigma_z,[\sigma_z,\rho(t)]] \notag\\
&=& -i \Theta^{\dagger} (-1) [ \bar{H} , \Theta \rho(t) \Theta^{\dagger}] \Theta 
-\gamma \Theta^{\dagger} [\sigma_z,[\sigma_z, \Theta \rho(t) \Theta^{\dagger} ]]\Theta \notag\\
&=& \Theta^{\dagger} {\cal L} \Theta \rho .
\end{eqnarray}
Note that $\Theta$ and $\Theta^{\dagger}$ are both antilinear. We then have $e^{-i {\cal L}_3 t} = \Theta^{\dagger} e^{-i {\cal L} t}  \Theta$ and 
$P'_3(t)= {\rm Tr} (\Pi_a e^{-i{\cal L}_3 t } \rho_0) ={\rm Tr} (\Pi_a e^{-i{\cal L} t } \Theta \rho_0 \Theta^{\dagger} )$. 
This leads to the conclusion that $P(t)$ is invariant under a sign change of $\bar{H}$ when either $\Theta \rho_0 \Theta^{\dagger}=\rho_0$ or $\rho_0$ is real. 

Combining the above analyses, we find that in the presence of Gaussian white noise $P(t)$ is still invariant under a simultaneous sign change of both $\nu$ and $\Delta$ when
 $\Upsilon\rho_0\Upsilon^{\dagger}=\sigma_z\rho_0\sigma_z^{\dagger}=\rho_0$ and $\rho_0$ is real.

\bibliography{noisyvaet}

\end{document}